# EXHAUSTIVE STUDY OF THREE-TIME PERIODS OF SOLAR ACTIVITY DUE TO SINGLE ACTIVE REGIONS: SUNSPOT, FLARE, CME, AND GEO-EFFECTIVENESS CHARACTERISTICS

**Shirsh Lata Soni*[1,2], Manohar Lal Yadav[2], Radhe Syam Gupta[1], Pyare lal Verma[1]**

[1]**Department of Physics, Awadhesh Pratap Singh University, Rewa, MP, India;**
[2]**KSKGRL, Indian Institute of Geomagnetism, Prayagraj, UP, India**
*sheersh171@gmail.com

**Abstract:**

In this paper, we present the multi-wavelength study of a high level of solar activity during which a single active region produced multiple flares/CMEs. According to the sunspot observations, the current solar cycle 24 manifest to be less intense in comparison with the previous recent sunspot cycles. In the course of the current sunspot cycle 24, several small and large sunspot groups have produced various moderate and intense flare/CME events. There are a few active regions with a large number of flaring activities passed across the visible disk of the Sun during 2012-2015. In this study, we consider the three periods 22-29 Oct 2013, 01-08 Nov 2013, and 25 Oct- 08-Nov 2014, during which 228 flares have been observed. Considering only active regions near the central part of the disk, 59 CMEs (halo or partial) have been reported among which only 39 events are associated with flares. We conclude that an active region with a larger area, more complex morphology and stronger magnetic field has a comparatively higher possibility of producing extremely fast CMEs (speed > 1500 km/sec). So that among the 5 X class flares of the reported periods, 3 of them (60%) are associated with a CME. The lift-off time for CME-flare associated events has a +15 to+30 minute time interval range after the occurrence time of associated flares suggesting that the flares produce the CMEs. Additionally, we compiled the geomagnetic storms occurring within 1-5 days after the CME onset. 10% of the 59 CMEs are related to a magnetic storm but all are moderate storms.

## 1. Introduction:

One of the most violent explosive phenomena in the solar atmosphere are Coronal Mass Ejections (CMEs). Active regions (ARs) are thought to be the most efficient producer of CMEs due to the accumulation of free energy in the photosphere. Solar active regions are the area with a strong concentration of magnetic flux with opposite polarities and appear as sunspot (dark patches) in white light on the photosphere. However, for the different active regions, the capability of generation CMEs could have different and it may not be necessary to produce CMEs from each active region. Various studies (Dhakal et al. 2019; Wang and Zhang, 2008) have shed light on the CME-producing capability of the active region from the last few decades but the relationship between the active region and CME is still unsolved. They have found a general trend that a larger, stronger, and more complex AR are more likely to produce a faster CME.

Although flares are different from CMEs, they are also a violent explosive phenomenon in the solar atmosphere. Flares can be classified into two, the confined ones and the eruptive ones according to whether or not they are associated with CMEs [e.g., Svestka and Cliver, 1992; Wang and Zhang, 2007; Schrijver, 2009; 2016]. Several (Webb & Howard 2012; Evans et al. 2013; Yashiro et al. 2006; Howard & Harrison 2013) studies have introduced a detailed study on the relationship between various parameters of CMEs (i.e. speed, acceleration, width, etc.) with flare characteristics (i.e. onset time, duration, class, etc. ). Although the idea of relation and association between CMEs and flare is well-argued and debated for many years. For this present work, we use CMEs from SOHO/LASCO (Solar and Heliospheric Observatory/Large Angle Spectro-metric Coronagraph) catalog because it contains a large set of CMEs with various parameters such as linear speed, width, acceleration, etc. The linear speeds are usually determined by the measurement of the height-time profile for the fastest-moving part of the CME front. The measured brightness appears to be surrounded by the entire solar disk that can be characterized as halo CME or in other words, the CME with measured angular width $360°$ is called a halo CME. The halo CMEs can be classified in three: disk halos (within 45 deg from disk center), limb halos (beyond 45 degrees but within 90 deg from disk center), and backside halo CMEs (Gopalswamy et al, 2008). It is well accepted that the front-side Halo CMEs (angular width $>120°$) are originated from the active regions located at $45°$ of the east to west visible solar disk (Yashiro 2013). The CMEs occurring close to the solar disk center are more are important to study as they are directed towards the Earth. Hence can influence and affect the Earth's magnetosphere and atmosphere more directly. Two primary requirements for the geoeffectiveness of CMEs are (1) the CMEs must arrive at Earth and (2) have a southward component of their magnetic field. CMEs originating from close to the disk center (within 45 deg from the disk center) propagate roughly along the Sun-Earth line, so the frontside halos are highly likely to arrive at Earth. We refer to them as disk CMEs. Frontside limb CMEs (originating at longitudes beyond 45 deg and up to 90 deg) propagate at an angle to the Sun-Earth line and only deliver a glancing blow to Earth's magnetosphere. CMEs ejected at angles exceeding 90 deg to the Sun-Earth line are unlikely to impact Earth.

This paper presents a study of the relationship between the active region and eruptive CMEs when the active region passes through the central region of the visible solar disk. As well as the association of halo/partial halos CMEs measured by coronagraph and solar flares measured in X-rays. We also describe the properties of halo CMEs such as speed, source location and associated flare size to investigate their effects on Earth's magnetosphere.

## 2. Data Sources:

In this presented work, we investigate the interplanetary consequences of halo/partial halo CMEs erupted from the same active region for a certain period. The CMEs considered are Halo and partial Halo events of width ≥130°. We take three active periods 22-29 Oct 2013, 01-08 Nov 2013, and 25 Oct- 08-Nov 2014 during 2012-2015 in the main phase of Solar cycle 24 listed in Table 1. Moreover, they are Earth-directed events that originated close to the center of the solar disk and propagate approximately along the Sun-Earth line. To investigate the source region of the CME, data from the Atmospheric Imaging Assembly (AIA; Lemen et al., 2012) onboard the Solar Dynamics Observatory (SDO; pesnell, Thompson, and Chamberlin, 2012) are used. We have analyzed EUV images of the Sun taken from 94 Å filter of AIA, which is sensitive to plasma at high temperature (~6 MK). The white light and magnetogram images from HMI provide a photospheric view of the active region. Each CME, the onset time, and initial speed have been estimated from the white-light images observed by the LASCO coronagraphs onboard the SOHO space mission. These CMEs and their storm details are obtained from the online catalogs http://cdaweb.gsfc.nasa.gov/istp_pub-lic/, www.lesia.obspm.fr/cesra/highlights/highlight07-html/ and http://wdc.kugi.kyoto.u.ac.jp/dst_realtime /index.html. In situ observations of ACE/Wind give details about the signatures of interplanetary CMEs and shocks and they are listed in the OMNI database. The arrival time of the IP shock of the four events is gathered from the OMNI website.

Here we summarize the details of eruptive events for all three selected active periods. In Table 1, we present all the eruptive events (CMEs/flares) corresponds with their source active region. Whereas in Table 2 shows only Halo and partial halo CMEs events (59) with their source location and basic parameters. We also present the details of associated flares, as we can see from the Table 2, that all CMEs are not associated with flare. We can also observe that the start time of flares is before the CME.

Table 1: Provides a list of all three periods 22-29 Oct 2013, 01-08 Nov 2013, and 25 Oct- 08-Nov 2014, which are considered for this study with CMEs and solar flares and accompanying data. The first column lists the considered period followed by the number of CME events in the second column. The third column lists the total no. of flare occurred in a particular period followed by the dominant active regions. In the last column, we have listed out the flares with their class and associated active region for a particular period of high activity.

| Period of high activity | CME events | Dominant active regions | Flare Record |
|---|---|---|---|
| 2013-oct 22-29 | 16 | 98 flares =11869 (1),11871 (1), 11873(2),11875 (57), 11877 (11), 11882 (21), 11884(1) | X class-03 (01-11875) (02-11882) |
| | | | M class- 23 (09-11875) (02-11877) (11-11882) (01-11884) |
| | | | C class-72 (01-11869) (01-11880) (01-11871) (02-11873) (09-11877) (50-11875) (08-11882) |
| 2013-nov 01-08 | 12 | 58 flares =11884(13), 11885(02), 11887(02), 11888(01), 11889(01), 11890(34), 11891(05) | X class- 01 (01-11890) |
| | | | M class- 07 (02-11884) (04-11890), (01-11891) |
| | | | C class-50 (11-11884), (02-11885), (02-11887), (01-11888), (01-1189), (29-11890), (04-11891) |
| 2014-oct 25 nov 08 | 31 | 72 flares = 12192(47), 12193(01), 12194(01), 12201(10), 12205(13) | X class- 03 (01-12205), (02-12192) |
| | | | M class- 19 (16-12192), (03-12205) |
| | | | C class-50 (29-12192), (01-12193), (01-12194), (10-12201) (09-12205) |

Table 2: In this list we summarize the same above mentioned the same period of activity [1] 22-29 Oct 2013, [2] 01-08 Nov 2013 and [3] 25 Oct- 08-Nov 2014, with CME and flare detail.

| | CME's detail | | | | | Flare's detail | | | | | |
|---|---|---|---|---|---|---|---|---|---|---|---|
| Date | Time | H/PH | Speed (km/s) | Width (deg) | AR | Location | Start time | Peak time | Last time | Flare class | |
| **2013 oct 22-29** | | | | | | | | | | | |
| 2013/10/22 | 04:24:05 | P Halo | 592 | 179 | 11875 | N07E16 | 04:12 | 04:21 | 04:25 | C4.0 | 4.0E-06 |
| 2013/10/22 | 21:48:06 | Halo | 459 | 360 | 11875 | N06E08 | 21:15 | 21:20 | 21:22 | M4.2 | 4.2E-05 |
| 2013/10/24 | 01:25:29 | Halo | 399 | 360 | 11875 | S10E08 | 00:20 | 00:28 | 00:48 | M9.3 | 9.3E-05 |
| 2013/10/25 | 08:12:05 | Halo | 587 | 360 | 11882 | S08E73 | 07:53 | 08:10 | 08:20 | X1.7 | 1.7E-04 |
| 2013/10/25 | 15:12:09 | Halo | 1081 | 360 | 11882 | S07E69 | 14:51 | 15:03 | 15:12 | X2.1 | 2.1E-04 |
| 2013/10/26 | 03:12:08 | P Halo | 473 | 208 | 11882 | S08E63 | 03:03 | 03:07 | 03:20 | C2.3 | 2.3E-06 |
| 2013/10/26 | 09:48:05 | P Halo | 460 | 141 | 11877 | S10E58 | 09:17 | 09:37 | 10:10 | M1.5 | 1.1E-03 |
| 2013/10/26 | 11:24:05 | Halo | 796 | 360 | 11882 | S10E56 | 10:11 | 11:08 | 12:12 | M1.8 | 1.8E-05 |
| 2013/10/26 | 19:12:05 | P Halo | 822 | 207 | 11877 | S07W28 | 19:24 | 19:31 | 19:38 | M2.0 | 2.0E-05 |
| 2013/10/27 | 12:12:06 | P Halo | 525 | 191 | 11882 | S09E65 | 12:33 | 12:38 | 12:46 | M3.5 | 3.5E-05 |
| 2013/10/27 | 18:12:05 | P Halo | 795 | 189 | 11875 | N12W91 | 17:48 | 17:53 | 18:04 | C9.1 | 9.1E-06 |
| 2013/10/28 | 02:24:05 | Halo | 695 | 360 | 11875 | N04W59 | 01:41 | 02:03 | 02:12 | X1.0 | 1.0E-04 |
| 2013/10/28 | 04:48:05 | P Halo | 1201 | 156 | 11875 | N08W71 | 04:32 | 04:39 | 04:53 | M5.1 | 5.1E-05 |
| 2013/10/28 | 15:36:05 | Halo | 812 | 360 | 11882 | S06E28 | 15:07 | 15:15 | 15:21 | M4.4 | 4.4E-05 |
| 2013/10/28 | 21:25:11 | P Halo | 771 | 142 | 11875 | N07W83 | 20:48 | 20:57 | 21:22 | M1.5 | 1.5E-05 |
| 2013/10/29 | 22:00:06 | Halo | 1001 | 360 | 11875 | N05W89 | 21:42 | 21:54 | 22:02 | X2.3 | 2.30E-04 |
| **2013-nov 01-08** | | | | | | | | | | | |
| 2013/11/01 | 15:48:05 | P Halo | 342 | 184 | 11890 | S11E02 | 15:17 | 15:22 | 15:25 | C1.8 | 1.8E-06 |
| 2013/11/02 | 04:48:05 | Halo | 828 | 360 | 11884 | S23W04 | 04:40 | 04:46 | 04:50 | C8.2 | 8.2E-06 |
| 2013/11/02 | 08:24:05 | P Halo | 798 | 177 | 11884 | S23W04 | | | | | |
| 2013/11/04 | 05:12:05 | Halo | 1040 | 360 | 11884 | S12W16 | 05:16 | 05:22 | 05:26 | M5.0 | 5.00E-05 |
| 2013/11/05 | 02:48:05 | P Halo | 512 | 126 | 11884 | S12W16 | | | | | |
| 2013/11/05 | 08:24:06 | P Halo | 850 | 197 | 11890 | S17E48 | 07:43 | 08:18 | 08:21 | M2.5 | 2.50E-05 |
| 2013/11/05 | 22:36:05 | P Halo | 562 | 195 | 11890 | S12E46 | 22:07 | 22:12 | 22:15 | X3.3 | 3.30E-04 |
| 2013/11/06 | 14:24:26 | P Halo | 347 | 122 | 11890 | S12E37 | 13:39 | 13:46 | 13:53 | M3.8 | 3.80E-05 |
| 2013/11/07 | 00:00:06 | Halo | 1033 | 360 | 11890 | S11W88 | 23:44 | 00:02 | 00:14 | M1.8 | 1.80E-05 |
| 2013/11/07 | 10:36:05 | Halo | 1405 | 360 | 11890 | S12E24 | 10:26 | 10:53 | 10:57 | C2.1 | 2.1E-06 |
| 2013/11/07 | 15:12:10 | Halo | 411 | 360 | 11890 | S12E21 | 15:39 | 15:47 | 15:49 | C1.6 | 1.60E-06 |
| 2013/11/08 | 03:24:07 | Halo | 497 | 360 | 11884 | S18W25 | | | | | |
| **2014-Oct25 Nov-08** | | | | | | | | | | | |
| 2014/10/25 | 21:36:05 | P Halo | 286 | 208 | 12192 | S15W45 | | | | | |
| 2014/10/26 | 19:36:05 | P Halo | 298 | 140 | 12192 | S15W45 | 19:59 | 20:21 | 20:45 | M2.4 | 2.40E-05 |
| 2014/10/27 | 06:00:05 | P Halo | 250 | 128 | 12192 | S13W45 | 05:21 | 05:38 | 05:47 | C4.9 | 4.90E-06 |
| 2014/10/27 | 06:48:05 | P Halo | 352 | 127 | 12192 | S13W44 | 06:56 | 07:01 | 07:07 | C9.6 | 9.60E-06 |
| 2014/10/28 | 20:24:05 | P Halo | 467 | 169 | 12192 | S13W44 | | | | | |
| 2014/10/29 | 12:36:05 | P Halo | 230 | 141 | 12192 | S13W44 | | | | | |
| 2014/10/30 | 23:24:05 | P Halo | 327 | 132 | 12192 | S13W44 | | | | | |
| 2014/10/31 | 09:12:08 | P Halo | 490 | 180 | 12201 | S05E57 | 09:19 | 09:23 | 09:27 | C2.0 | 2.00E-06 |
| 2014/10/31 | 18:22:32 | P Halo | 242 | 120 | 12201 | S05E57 | | | | | |
| 2014/11/01 | 05:00:05 | P Halo | 1628 | 159 | 12201 | S22E52 | 04:44 | 05:34 | 07:05 | C2.7 | 2.70E-06 |
| 2014/11/01 | 06:00:06 | P Halo | 740 | 160 | 12201 | S22E52 | 04:44 | 05:34 | 07:05 | C2.7 | 2.70E-06 |
| 2014/11/01 | 15:36:05 | P Halo | 647 | 218 | 12201 | S05E45 | | | | | |
| 2014/11/01 | 19:12:05 | P Halo | 659 | 216 | 12201 | S05E41 | | | | | |
| 2014/11/02 | 10:00:06 | P Halo | 461 | 168 | 12201 | S05E36 | | | | | |
| 2014/11/02 | 21:24:05 | P Halo | 289 | 180 | 12201 | N15E89 | 20:44 | 21:05 | 21:25 | C9.4 | 9.40E-06 |
| 2014/11/03 | 11:36:06 | P Halo | 225 | 121 | 12205 | N17E89 | 11:23 | 11:53 | 12:17 | M2.2 | 2.2E-05 |
| 2014/11/03 | 12:00:05 | P Halo | 447 | 196 | 12201 | S05E11 | | | | | |
| 2014/11/03 | 15:12:09 | P Halo | 226 | 148 | 12205 | N17E89 | | | | | |
| 2014/11/03 | 18:58:05 | P Halo | 459 | 133 | 12205 | N17E89 | 19:10 | 19:16 | 19:21 | C1.7 | 1.70E-06 |
| 2014/11/03 | 23:12:33 | P Halo | 638 | 155 | 12205 | N17E89 | | | | | |
| 2014/11/04 | 06:12:05 | P Halo | 190 | 143 | 12205 | N15E82 | | | | | |
| 2014/11/04 | 08:48:06 | P Halo | 627 | 175 | 12205 | N15E82 | 08:42 | 09:18 | 09:35 | M2.3 | 2.3E-05 |
| 2014/11/04 | 21:28:11 | P Halo | 211 | 149 | 12205 | N15E82 | | | | | |
| 2014/11/05 | 03:12:08 | P Halo | 198 | 144 | 12205 | N15E82 | | | | | |
| 2014/11/05 | 08:12:05 | P Halo | 247 | 172 | 12205 | N17E76 | | | | | |
| 2014/11/05 | 10:00:05 | P Halo | 386 | 182 | 12205 | N20E68 | 09:26 | 09:47 | 09:55 | M7.9 | 7.9E-05 |
| 2014/11/05 | 19:48:05 | P Halo | 608 | 203 | 12205 | N17E65 | 18:50 | 19:44 | 20:15 | M2.9 | 2.90E-05 |
| 2014/11/06 | 04:00:05 | P Halo | 641 | 210 | 12205 | N17E58 | 03:29 | 03:50 | 05:12 | M5.4 | 5.40E-05 |
| 2014/11/06 | 06:12:05 | P Halo | 320 | 130 | 12205 | N17E58 | | | | | |
| 2014/11/07 | 18:08:34 | P Halo | 795 | 293 | 12205 | N13E35 | | | | | |
| 2014/11/08 | 16:36:05 | P Halo | 426 | 141 | 12205 | N15E23 | 16:44 | 17:55 | 18:21 | C4.4 | 4.40E-06 |

## 3. Observations:

To identify and study the early evolution of active regions on solar disk we use a multi-wavelength view of source regions of the reported active period. For the present study, we choose three reported Active periods from 2012 to 2015. In Table 1, we present the NOAA active regions that appeared in a particular period with their location in the heliographic disk. In Table 1, the first column shows the reported active periods, the second column shows the total number of CME event occurred during the particular duration, in the third column shows the total number of flare erupted from various active region (i.e: 11869 (1), where the 11869 is the active region and (1) is noted for the number of flares), in the last column of we reported the flare with their class and source active region (i.e.: '02-11884' where 02 is the number of flares (X, M or C class) and 11884 is the source active region). We also mentioned the first and last appearance of all active regions, those appeared on selected active periods (Table 2). As we are interested only those NOAA active regions produce multiple Halo and partial Halo Coronal Mass Ejection while transmitting from close to the center of visible solar disk. So the selected active regions are as follows:

**First Active period: 22-29 October 2013:** During this period of high activity, three active regions AR-11875, AR-11877, AR-11882 are identified which produce multiple Halo and partial Halo CMEs. Active region NOAA 11875 appeared on the solar disk on 18 Oct 2013 at the heliographic location N07E55 with simple alpha type magnetic configuration the active region grew rapidly in term of size and as well magnetic complexity and turned in to beta-gamma –delta category on 22 Oct 2013. And AR-11875 disappear on 30 Oct 2013. The second active region we recognized here is AR-11877, which appeared on 19 Oct 2013 at location S11E61 on solar disk with alpha-type magnetic configuration. Then it grows into alpha-beta-gamma category on 24 Oct 2013 and finally disappears on 31 Oct 2013 from the heliographic solar disk. Similarly, AR-11882 first appeared in the heliographic disk on 25 Oct 2013 at the location S08E59 and after about 5 days disappeared on 06 Nov 2013. During their existence on the photosphere, it erupts 21 flares (02-X class; 11-M class and 08- C-class).

**Second Active period: 01-08 November 2013:** During this period of high activity, two active regions AR-11884 and AR-11890 is identified which produces multiple Halo and partial Halo CMEs. Active region NOAA 11884 appeared on the solar disk on 27 Oct 2013 at the heliographic location S09E65 with simple alpha type magnetic configuration the active region grew rapidly in term of size and as well as magnetic complexity and turned in to beta-gamma-delta category on 28 Oct 2013 and after ~11 days of existence disappeared from disk on 08 Nov 2013, during its life it erupts multiple Halo/P Halo CMEs and flares (02-M class; 11-C class). The second active region we recognized here is AR-11890, the first appearance of this AR on 03 Nov 2013 at location S09E61 on solar disk with alpha-type magnetic configuration. Then it grows into beta-gamma-delta category on 05 Nov 2013. The transit of this active region from the center of the solar disk was ~6 days and disappear on 14 Nov 2013 from the heliographic solar disk. This AR was so eruptive that it erupts 34 flares with multiple intensities (01- X class; 004- M class; 29- C class).

**Third Active period: 25 October – 08 November 2014:** During this period of high activity, three active regions AR-12192, AR-12201, AR-12205 are identified which produce multiple Halo and partial Halo CMEs and several flares. Active region NOAA 12192 appeared on the solar disk on 18 Oct 2014 at the heliographic location S13E54 with simple alpha type magnetic configuration the active region grew rapidly in term of size and as well as magnetic complexity and turned in to beta-gamma-delta category on 22 Oct 2014. And AR-12192 disappear on 31 Oct 2014 after 13days. Due to its highly complex magnetic configuration, it erupts 02-X class, 16-M, class and 29-C class flares. The second active-region we recognized here is AR-12201, which appeared on 30 Oct 2014 at location S05E53 on solar disk with alpha type magnetic configuration. Then it grows into beta category on 09 Nov 2014 and finally disappears on 09 Nov 2014 from the heliographic solar disk. On 05 Nov 2014 from heliographic location N15E53, the third AR-12205 appeared. During its existence on the photosphere, it erupts a total of 13 solar flares (01- X class; 03-M class and 09-C Class flares).

### 3.1 Multi-wavelength view of solar active region

In the photosphere, the solar active regions consisting of the distribution of opposite magnetic polarities of sunspots during the occurrence of solar flare/CMEs. These active regions are covered by large overlying coronal loops. It is accepted that the onset of solar eruptions is related to some typical and complex magnetic structures. The introduction of scanning magnetographs makes it easy to map the full disk of the Sun to explore the complete view of the magnetic morphology of the sun surface from the magnetogram, which shows the bipolar nature of ARs. The solar and heliospheric activities reported in this paper correspond to the major eruptive activities for three periods. As above mentioned these particular most active durations, first period (22-29 Oct 2013) for active region NOAA AR -11875, AR-11877, AR-11882 and second period (01-08 Nov 2013) for AR NOAA-11882; NOAA AR-11884 and AR-11890; third period (25Oct-08Nov 2014) for NOAA AR-12192, AR-12201 and AR-12205. These are the active regions that transmitting from the mean location of the heliographic solar disk. To compare the photospheric structure of the active regions with the overlying chromospheric and coronal layers, we show HMI magnetogram and AIA 94 Å images of the active regions in panel (a) (b) and (c) for all three selected active periods respectively. From these HMI magnetograms and AIA 94 Å images, we can see the heliospheric position of selected active regions for reported Active periods. The HMI magnetograms, showing the distribution of sunspot and their magnetic polarity. White and black color in the HMI magnetogram indicates the positive and negative magnetic polarity region respectively, in the photosphere.

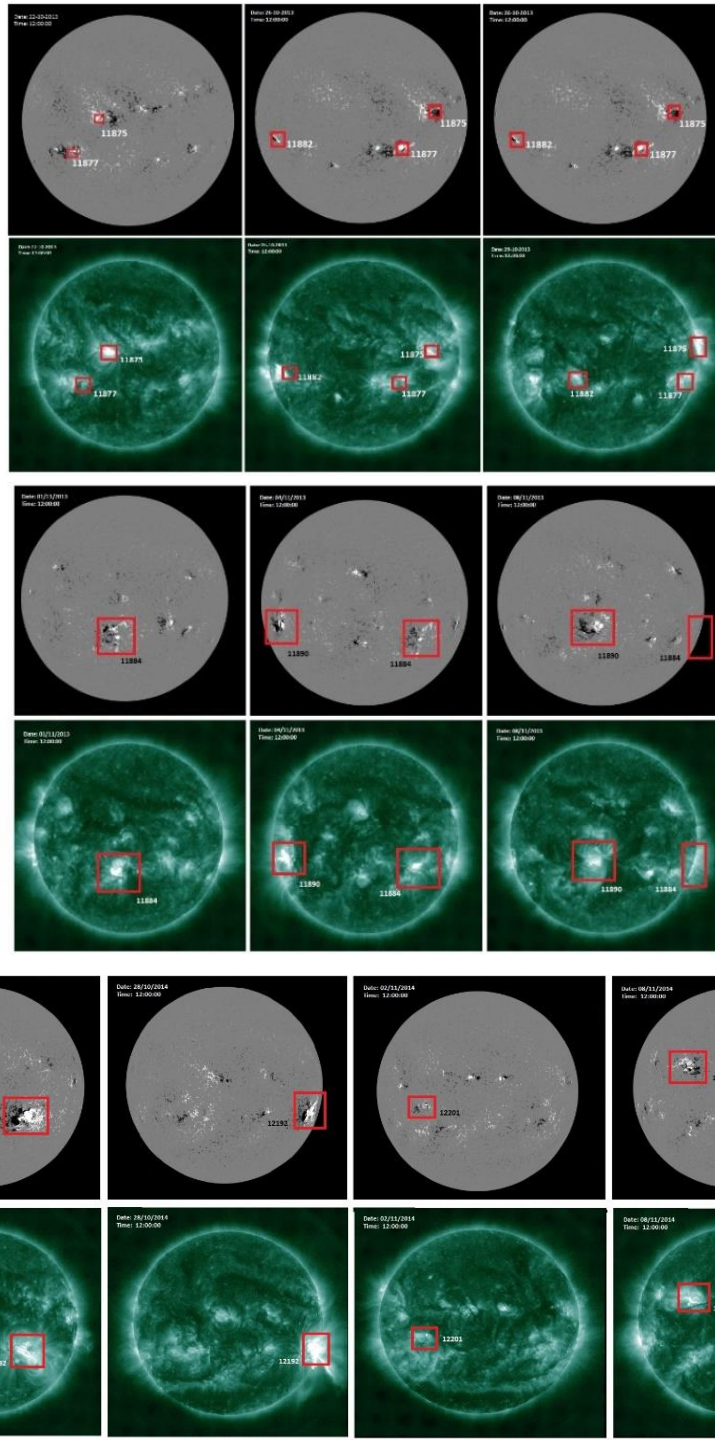

Figure 1: Multi-wavelength view of Active regions showing the location where sunspot occurred during selected active periods. (a) 22-29 Oct 2013, (b) 01-08 Nov 2013 and (c) 25 Oct- 08-Nov 2014. In all three panels, the above row of images is the HMI magnetograms, showing the distribution of sunspot and their magnetic polarity. White and black color in HMI magnetogram indicates the positive and negative magnetic polarity region respectively, in the photosphere. In panels, the below row of images is AIA 94 image of active regions showing hot coronal loops.

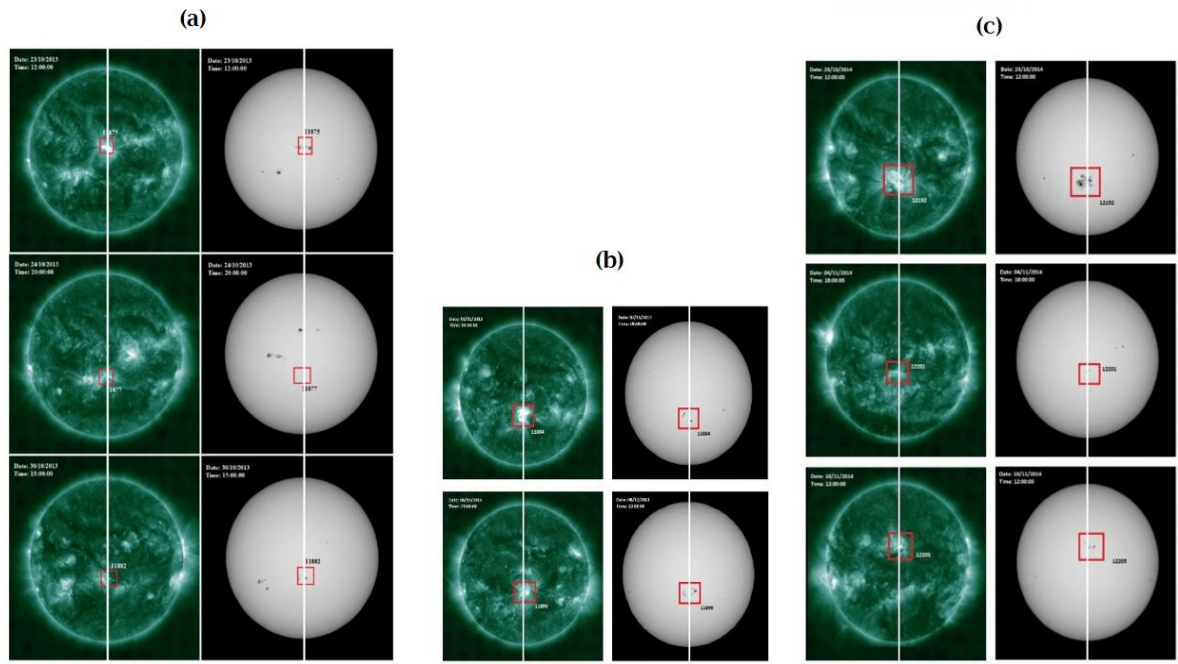

Figure: 2 These solar disks show the selected active region passes through the central meridian of the heliographic disk. In the first column we shown solar disk in 94 Å band and in the second column in the figure we choose the continuum image of the solar disk (a) Shows the first Active period with the three most active region AR 11875, AR 11877, and AR 11882 dated 23/10/2013, 24/10/2013, and 30/10/2013 respectively. (b) Shows the first Active period with the three most active region AR 11884, AR 11890 dated 02/11/2013, and 04/11/2013 respectively. (c) Shows the first Active period with three most active region AR 12192, AR 12201, and AR 12205 dated 23/10/2014, 04/11/2014, and 10/11/2014 respectively.

In figure 2 we show the selected active regions for all three reported active periods are passing through the central meridian of heliographic disk. For comparison of the photospheric structure of the sunspot region here we choose AIA 94 Å band and HMI continuum images, which is shown in first and second column of figure 2. From figure 2(a) for first Active period 22-29 Oct 2013, we can see that NOAA AR-11875 on 23 Oct 2013 12:00:00UT; NOAA AR-11877 on 24 Oct 2013 20:00:00UT and NOAA AR-11882 at 30 Oct 2013 12:00:00UT are passing through the central meridian. From figure 2(b) we can see the NOAA AR-11884 at 02 Nov 2013 00:00:00 UT and NOAA AR-11890 on 08 Nov 2013 23:00:00UT are passing through the central meridian during the active period 01-08 Nov 2013. Similarly, in figure 2(c) for period 25Oct-08Nov 2014, NOAA AR-12192 at 23 Nov 2014 12:00:00 UT; NOAA AR-12201 on 04 Nov 2014 18:00:00 UT; NOAA AR-12205 at 10 Nov 2014 12:00:00 UT crossed the central meridian of heliographic the solar disk.

## 3.2 Early evolution of Active regions:

The magnetic classification of the active region describes the configuration of the magnetic flux and the sunspots. The Mount Wilson (Hale, et.al 1919) classification system for the sunspot group is a very simple way of classification. In this classification, the active region contains a single sunspot having single (same) polarity known as alpha. The next classification is for the active region having two the sunspots (the sunspot group) of opposite polarity, called beta. Classification gamma indicates at AR has a complex region of the sunspot with intermixed polarity. The delta classification assigned to at least one the sunspot in the region contains opposite magnetic polarities inside of a common penumbra separated by no more than 2° in heliographic distance (24 Mm or 3300 at disk center). Several researchers confirm by statistical studies that the active region contained delta type magnetic configuration are more flare productive (Sammis, Ternullo et al. (2006), Tang, & Zirin (2000), and Guo, Lin, & Deng (2014)). Although the flare productivity from active regions also correlated with other properties such as active region lifetime duration, the sunspot area, total magnetic flux and various other parameters.

As we can see from the above section (multi-wavelength view), that active regions have changed in their size, area, and magnetic configuration. In this section of observation, we present the growth of active region concerning to area (size) with time and magnetic complexity as well. The area of active region plotted from the first appearance to the last appearance of active region on the solar disk. In plots (Figure 3), the shaded part shows the duration of reported active periods. From the plots, we can see that active regions grew rapidly in terms of area as well as in magnetic configuration. At the starting of evolution the area is very less (~100 to 200 msh) with simple alpha or beta type magnetic configuration. As they grow, they turned into the complex the beta-gamma-delta category of magnetic configuration. During the period, when the region with beta-gamma-delta magnetic configuration category, the number of flare events is comparatively high. We have also mentioned the number of eruptive flares on particular day with their intensity class. For example, in first graph, the active region 11875 have beta-gamma-delta magnetic configuration on 28 Oct 2013, inside the bar of 28- Oct- 2013 we show the number of flares C-2; M-3; X-1 (where C stand for C class flare; M is for M-class flare, and X is for X-class flare). As we can observe from the plots that active region having more complex magnetic configuration, erupt a greater number of flares.

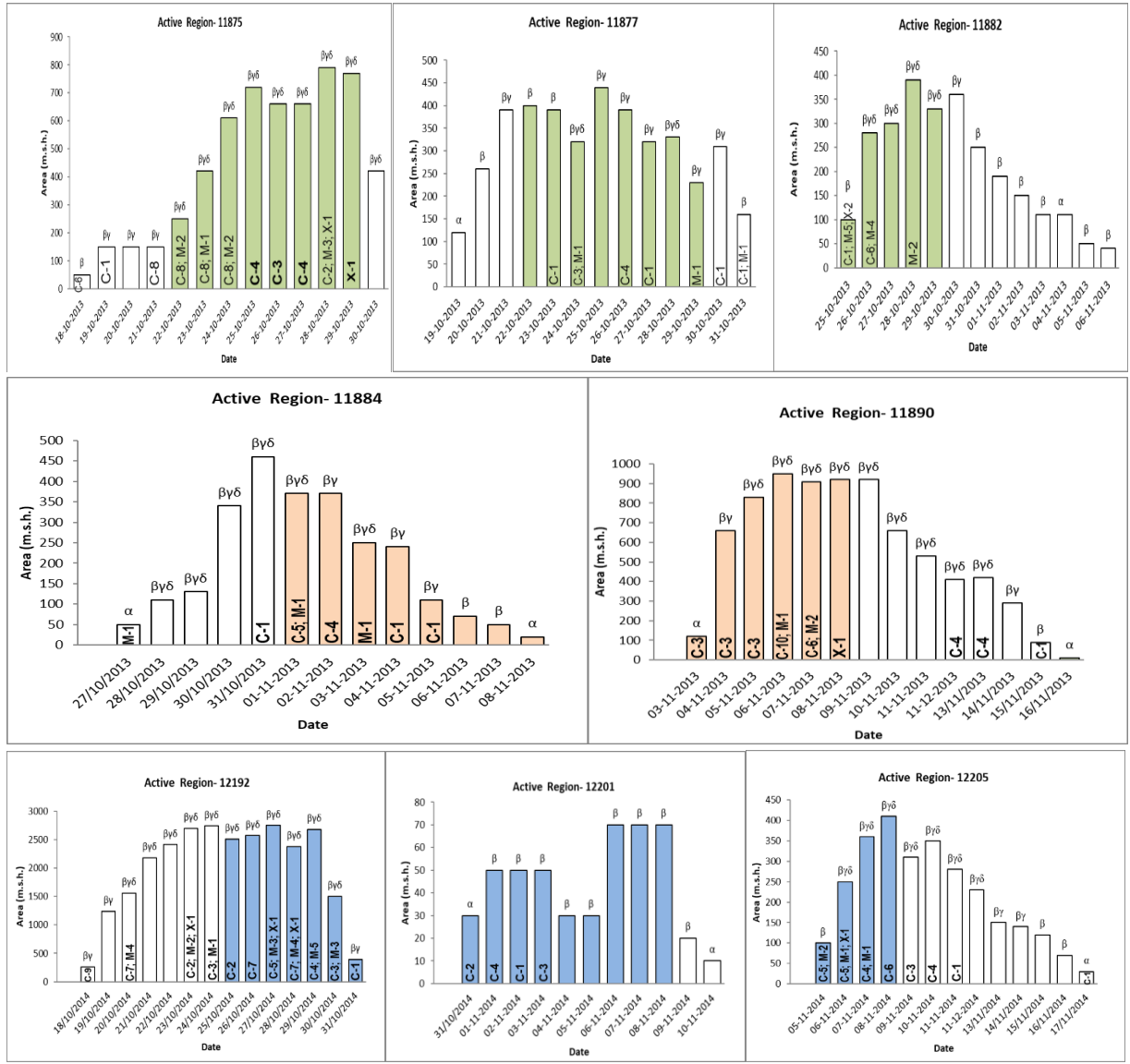

Figure 3: bar plots showing the evolution of active region in magnetice complexcity and area concerning the time. Here the topmost three bar plots for the observed active regions for the first duration (22-29 Oct 2013). The middle two plots are for the active regions observed during second duration (01-08 Nov 2013). The below three bar plots for the active regions occurred on third duration (25 Oct- 08 Nov 2014). The shaded part (Green- 22-29 Oct 2013; Orange- 01-08 Nov 2013 and Blue- 25Oct-08-Nov 2014) shows the reported period.

### 3.3 CME and Flare association:

Coronal mass ejections (CMEs) and the solar flares are the most energetic phenomena in our solar system. A CME can be defined as changes observed in the coronal structure which occurs on time scales of a few minutes to several hours. The outward motion can be observed and involves the instance of a newly-appearing, bright, white-light features can be seen as being discrete in coronagraph imagery. We have observed the flare events associated with CMEs for the selected active duration. For all three reported active period, total 59 Halo/partial Halo CMEs (16 CMEs from 22-29 Oct 2013; 12 CMEs from 01-08 Nov 2013 and 29 CMEs from 25 Oct-08 Nov 2014) are observed. Among them only 39 CME events were associated with flares. In this CME-flare associated events 15 flares are C-class, 19 flares are M-class and only 05 CME events are associated with X-class flare which is presented by GOES X-ray plots in Figure 4. Several researchers conclude three basic ideas related to CME-flare relationship such as: according to Dryer 1996: the flares produce CMEs, according to Hundhausen, 1999: flares are by products of CMEs; and according to Harrison 1995; Zhang et al., 2001: flares and CMEs are part of the same magnetic eruption process.

To select associated flare with specified CME event, we used two conditions, firstly the special condition and second is temporal condition. According to special conditions, the CME and flare should be from the same location and active region as well. And for the temporal condition required the CME eruption time is must be simultaneous to the start time of flare.
Figure 5 shows the event frequency based on their speed. From Figure we can see that most of the CME events have speed in the range of 200- 400 km/sec. even very few CMEs have high speed (>1200 km/sec).

We have picked up 59 CMEs from all reported active regions. We found 39 CMEs associated with flare. Figure 6 shows a histogram for the time delay of the flare and associated CME. Here we are taking spatial and temporal conditions into account. I have picked up all the CMEs that occurred between ±1 hour to the start time of their associated flare. From the observation, we found weak co-relation between soft X-ray flare intensity and CME speed. We also found weak correlation between peak time, decay time and total duration with associated CME speed.

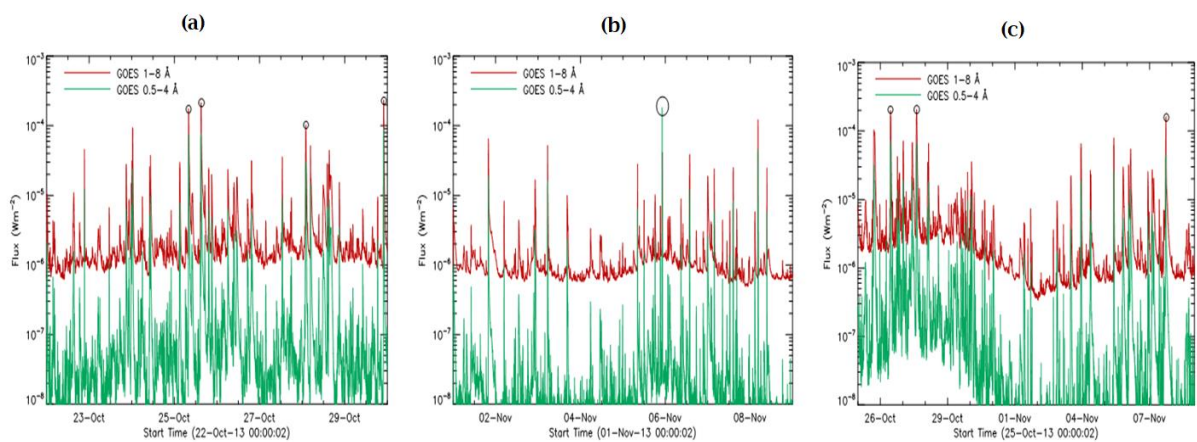

Figure 4: GOES X-ray plots in two wavelength bands (red) 1-8 A and (green) 0.5-4 A for duration 22-29 Oct 2013 (a), period 01-08 Nov 2013 (b), and third period 25Oct-08Nov 2014 (c). The circle indicates the X-class flares.

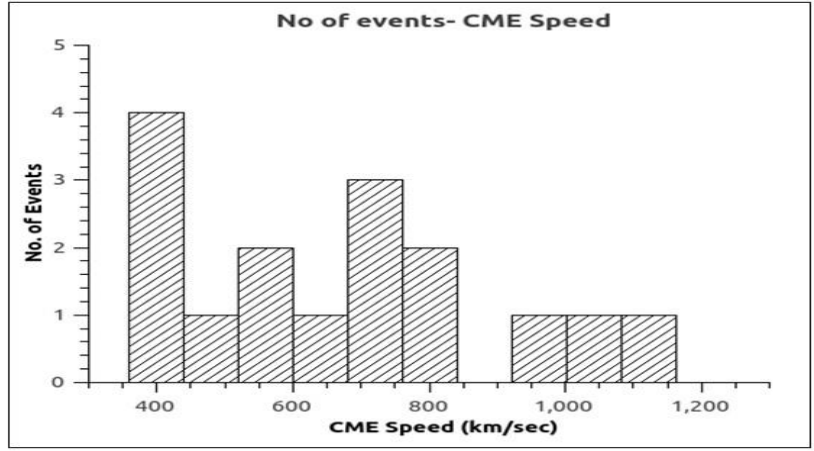

Figure 5: Histogram for the number of CME events according to their speed.

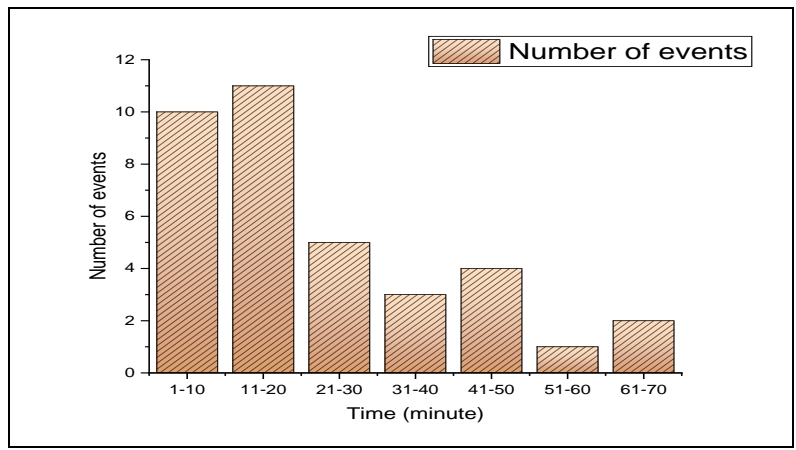

Figure 6: histogram for CME-flare time duration. Detection of time delay in CME with associated solar flare.

To study the effect of CME-lift of time on the dependence of CME-flare relationship, we present total CME and their associated flare events in two groups concerning to the start time of flare as the reference level in Figure 7. Here the first group is CME time delay events and another one is CME time over events. We have reported total of 59 CMEs in which only 39 CMEs are associated with flares. We have found that 54% of CME-flare associated events (32 events) are CME time over events and a few numbers (06 events) of CME-flare associated events in the group of CME time delay events. These results uphold the idea that the flares develop the CMEs as postulated by Dryer et al., 1996 instead of the idea that flares are an appendage of as suggested by Hundhausen, 1999.

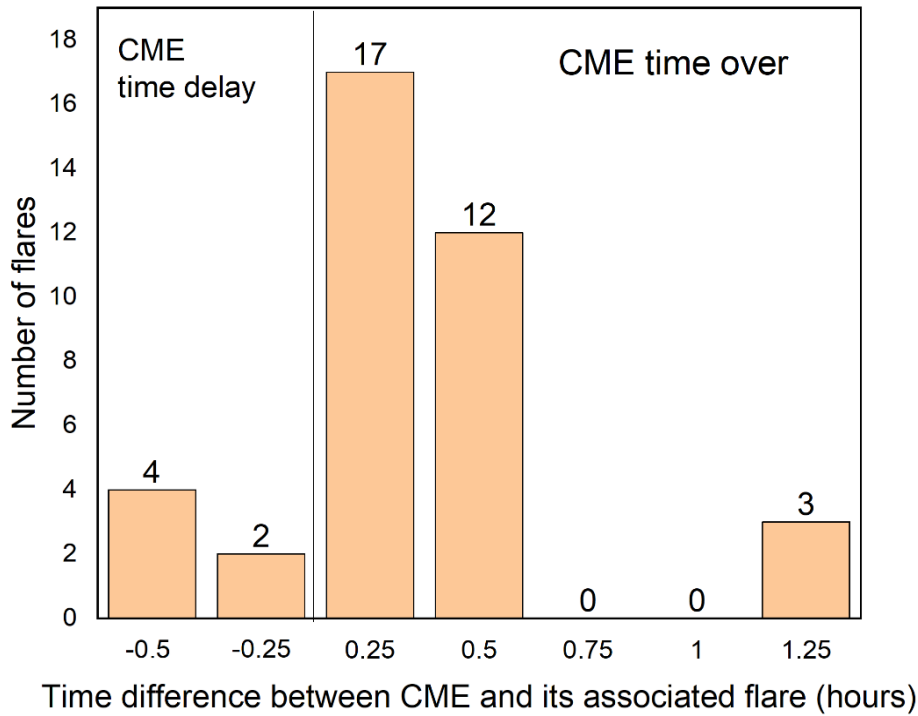

Figure 7: Histogram of the time interval between CME first appearance and start time of the associated flare.

### 3.4 Geo-effectiveness of CMEs/flares:

It is well established that the halo or partial halo CMEs are likely to affect the Earth's magnetosphere and generate geomagnetic storms (Richardson & Cane, 2010, Schmieder et al., 2020). The ability of CMEs to produce geomagnetic storms is known as geo-effectiveness, it is measured in terms of the geomagnetic index as the 'disturbances storm time' or Dst index. According to Loewe and Prolss [1997], based on the minimum value of the Dst, geomagnetic storms can be classified into five groups: weak (30 to 50 nT), moderate (50 to 100 nT), strong (100 to 200 nT), severe (200 to 350 nT), and great (< 350 nT). The strong, severe, and great storms are all caused by CMEs. However Weak and moderate storms could be caused by both CMEs and co-rotating interaction regions (CIRs) (Gosling et al., 1990). Primary requirements for the geo-effectiveness of CMEs are (1) the CME must be reach at Earth and (2) having southward component of their magnetic field (Soni et al., 2019). To determine the geo-magnetic effects of CMEs, we consider the minimum Dst values 1-5 days after the CME onsets. In Table 3, we present the interplanetary shock arrival time and their speed followed by Dst values corresponding to the associated CME and flare with their source active region. As we have selected 59 CME events, but here we observe only 06 Dst events, all these observed geomagnetic storms are moderate (-50 nT > Dst > -100 nT). It is clear that not all the earth directed CMEs can produce geomagnetic storms. The structure of magnetic field with the CME may determine the possibility of occurance of a geomagnetic storm (Burton et al. 1975; Cane et al. 2000; Shirsh et al. 2020). Here the associated CMEs have a speed range from ~200 -1050 km/sec, it shows that the speed of CMEs has weak correlation with the intensity of the geomagnetic storm.

Table 3: shows the geo-effectiveness of CMEs errupted from dominant active regions of selected durations. Table summarizes the arrival of shock and their characteristics at 1 AU.

| Active Region | CME Date & Time | Speed (km/sec) | Halo/P Halo | Flare | IP shock Date & Time | IP shock Speed (km/sec) | Dst (in nT) | Bz (in nT) |
|---|---|---|---|---|---|---|---|---|
| 2013-Oct 22-29 | | | | | | | | |
| 11875 | 28-10-2013 02:24 | 636 | Halo | X 1.0 | 29-10-2013 09:33 | 343 | -59 | -8 |
| 2013-nov 01-08 | | | | | | | | |
| 11884 | 04-11-2013 05:12 | 1040 | Halo | M 5.0 | 06-11-2013 13:37 | 329 | -52 | -14 |
| 11890 | 07-11-2013 00:00 | 1033 | Halo | M 1.8 | 08-11-2013 22:00 | 420 | -79 | -12 |
| 11884 | 08-11-2013 03:24 | 497 | Halo | -- | 11-11-2013 10:31 | 480 | -68 | -8 |
| 2014-oct 25 nov 08 | | | | | | | | |
| 12192 | 25-10-2014 21:36 | 236 | P Halo | -- | 27-10-2014 02:00 | | -58 | -7 |
| 12205 | 09-11-2014 10:24 | 633 | P Halo | -- | 11-11-2014 07:00 | 480 | -63 | -9 |

## 4. Discussion and conclusion:

As we know that the solar cycle 24 is less active in comparison to recent previous cycles. In the course of current the sunspot cycle, several small and large the sunspot groups have produced moderately intense flare/CME events. From above study, we can see that the active region which passes through the central part of hemispheric the solar disk may most probably produce earth directed Coronal Mass Ejections. In Table 1, we mentioned the active regions observe in reported duration which produces multiple classes of flares. For this present work we try to identify the phases of moderate to the high level of the solar activity. So here we consider few particular time durations which are most active and erupted number of flare/CME events and give rise to intense/moderate geomagnetic storms. We also examine the CME-flare relation and identify the effects of these CMEs into interplanetary space and near-Earth environment. Here we report three periods (approximately 08-15 days) (1) 22-29 Oct 2013, (2) 01-08 Nov 2013, (3) 25Oct-08 Nov 2014. From reported active periods we obtained total of 59 CME events. Among these 59 CMEs only 39 CMEs are associated with the solar flare. And only a few (06) CMEs are geo-effective, which produces moderate geomagnetic storms.

From the above study we conclude the following points:

- As the results of the above study, we can claim that the size, complexity and strength of active regions have little correlation with the kinematic properties of CMEs, but have significant effects on the productivity of CMEs.
- Our study reveals and justifies the previous results [Wang and Zhang, 2008] that CME speed is dependent on the area of active region, magnetic complexity and flux, as a trend, an active region with larger area, more complex morphology and stronger magnetic field has a comparatively higher possibility of producing extremely fast CMEs (speed > 1500 km/sec). However, a weak correlation was found between the active region parameters (the area magnetic configuration and magnetic flux) and CME width.
- Our analysis verifies the previous results (Waldmeier, 1955; Choudhary et al.,2013) that flare is the most intense and numerous during the flux reaches its maximum value and flare numbers peak when the sunspot area of the active region is at maximum.
- The active region which passes through the central part of the solar would most probably produce Earth directed (Halo CMEs) coronal mass ejection.
- The lift-off time for CME-flare associated events having a +15 to +30 minutes time interval range after the occurrence time of associated flares.
- We found that the CME-flare associated rate for reported events is dominant for X-class flares.
- We have found that 54% of CME-flare associated events (32 events) are CME time over events and a few number (06 events) of CME-flare associated events in the group of CME time delay events. These results uphold the idea that the flares develop the CMEs as postulated by Dryer et al., 1996 instead of the idea that flares are appendage of as suggested by Hundhausen, 1999.
- We observe only few (06) moderate geomagnetic storm events correspond with ~59 Halo/ P Halo CMEs having speed ~200 -1050 km/sec implies that not all earth directed CMEs produce geomagnetic storm and weak correlation between CME speed and intensity of Dst index.

**Acknowledgment and Data availability:**

Authors acknowledge to NASA's open data policy for SOHO, STEREO, SDO and WIND data. We acknowledge the NOAA/NGDC for making the GOES soft X-ray and Dst data available to be included in the CME catalog of LASCO/SOHO available at NASA's CDAW data warehouse https://cdaw.gsfc.nasa.gov/CME_list/. We are gratefully acknowledge support and advice from Dr, Bhuwan Joshi Sir (Scientist, Udaipur Solar Observatory, PRL, Udaipur). The confirmation of Associated phenomena with reported CME such as the solar flare, radio burst, interplanetary disturbances and geo-effectiveness analysed using observation available from various space and ground based instruments and data are available taken from https://www.ngdc.noaa.gov/stp/spaceweather/the solar-data/the solar-features/the solar-flares/x-rays/goes/xrs/;http://www.e-callisto.org/;http://wdc.kugi.kyoto-u.ac.jp/wdc/aedstcited.html; https://omniweb.gsfc.nasa.gov/form/dx1.html etc. All authors have no conflicts of interest.